# DAQ meta-software for HEP experimental setups

(preprint)

S. Ryzhikov
NRC Kurchatov Institute - IHEP, Protvino, Russia
E-mail: ryzhikov@ihep.ru

**Abstract:** Meta-software for data acquisition (DAQ) is a new approach to design the DAQ systems for experimental setups in experiments in high energy physics (HEP). It abstracts from experiment-specific data processing logic, but reflects it through configuration. It is also intended to substitute highly integrated DAQ software for a swarm of single-functional components, orchestrated by universal meta-software.

# Contents

# 1   Introduction

The crew of a large-scale high energy physics (HEP) experiment often has to deal with an inevitably complex data acquisition (DAQ) process. Modern experiments combine unique hardware with sophisticated networking and computing facilities for data handling and logging. This process is monitored and supervised by DAQ software, which represents the structure of the entire DAQ chain. The design of this software can affect the data-collection efficiency, and thus, considerable effort is usually made to optimize it.

Almost all major experiments develop their own DAQ software, mainly because it incorporates a significant amount of detector-specific code that cannot be reused in other experiments. The development of such software can be quite costly. An alternative approach is to use DAQ software frameworks, as artdaq developed in Fermilab [1]. However, general-purpose DAQ frameworks have certain limitations and often do not match the specific needs of a large experiment.

We propose a novel approach to build an entire DAQ system around what we call meta-software. "Meta" refers to the provision of only high-level functionality and to the abstraction from the actual data handling. In this approach, the complete DAQ chain of an experiment is described by the meta-software configuration, which implies that the software does not include any detector-specific code. Hence, several HEP experiments can participate in development and use each other's achievements without modifications. Experiment-specific data processing, of course, has to be implemented, but in separate software units that should interact asynchronously with meta-software.

In order to test this new approach, we implemented our proof-of-concept version of DAQ meta-software, ηDAQ. Although it is not intended to be used in any large-scale experiment yet, it can serve as a reference point for the development of similar systems.

## 1.1   Software in HEP experiment

Any type of software in the HEP experiment can be attributed to one of the three areas: detector control system (DCS), DAQ, or data analysis (Fig. 1). DCS software is usually developed by creators of each detector independently, and in most cases, intended for experts rather than for shifters. Here, we will consider only the DAQ area, which is operated by a crew.

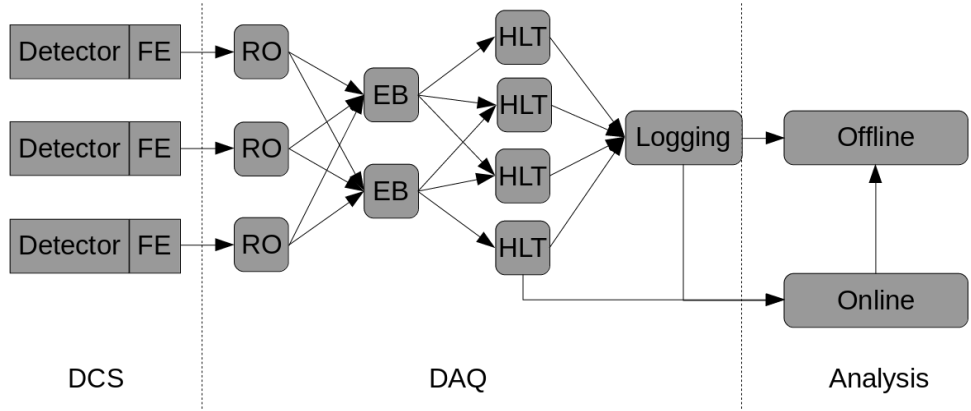

*Figure 1 General DAQ scheme of a modern HEP experiment.*

DAQ software functionality can be divided into three layers:
- **Data processing** includes readout (RO) for each detector, event building (EB), high-level triggering/filtering (HLT) and data logging. It is strongly coupled with the hardware.
- **Supplementary functionality** covers interaction with other systems and usually includes metadata index, hardware configuration utilities, run control logic, user interface, monitoring, and storage management. It is basically the same for all the HEP experiments, but each experiment usually implements it in an individual way.
- **Advanced functionality** provides utilities and automation tools, such as data quality assessment, expert system for shifters, and autopilot with elements of artificial intelligence.

## 1.2 DAQ Frameworks vs. Meta-Software

DAQ frameworks provide off-the-shelf supplementary functionality and templates for data processing code. For a working DAQ, each experiment needs to develop the missing parts of the data processing code in accordance with the specification of the framework. Adoption of framework mandates agreement with all the decisions of the framework authors such as programming language, data format, operational logic.

As with the frameworks, meta-software also provides off-the-shelf supplementary functionality. However, the data processing parts have no restrictions on the way they could be implemented. For their interaction with meta-software, it is sufficient even if they provide only notifications about changes in their state. For example, each instance of event-builder should notify meta-software of the unique identifiers of processed data chunks.

Today, the data processing functionality can be implemented not only in software, but also in FPGA-based hardware. For example, instead of waiting for readout, front-end electronics are now able to continuously stream the data to the event builders via TCP[1] protocol [2]. Event-building and some HLT algorithms also could be offloaded to FPGA-based hardware. For the meta-software, it does not matter how the actual data processing is organized.

---

[1] Or any reliable UDP-based protocol.

## 2 ηDAQ

In ηDAQ, our implementation of DAQ meta-software, we exploit the fact that in modern HEP experimental setups, all the data processing is performed asynchronously, since it has no real-time control loops and data fragments are already time-stamped by the frontend electronics.

We implemented our meta-software as a set of web-applications, which interact with each other via simple web service API. As implementation simplicity was the main goal, we have chosen Python and the Flask - SqlAlchemy - Bootstrap 4 - Vue.js software stack, which is popular among developers of web-based applications. Our implementation consists of four microservices, which are briefly described in the following sections.

### 2.1 Configuration and metadata

This component is a central database for all Run metadata and the index of all acquired data chunks. Any component can request the current or past Run configuration values via a simple API call.

We have "recycled" the commonly used but outdated concept of *Run* as a data recording session. This concept originated in the era of magnetic tapes and is still used in almost every HEP experiment. With modern data warehouses, unclaimed data can be deleted later at no additional cost. Thus, we have a strong belief that manual interruption of the data recording is not required anymore.

Our *Runs* are periods of data collection with certain conditions. If conditions were fulfilled during the given time slot, the corresponding chunk of recorded data is considered to be valid for this *Run*. These conditions are presented as valid variable values. For numerical variables, units, tolerance, and range can be specified.

For example, a part of "cosmic" *Run* configuration for ECAL detector may look like this:

```
ecal1.state = "RUNNING"
ecal1.hv = (2375.0) V ±2.0 (0, 3100)
trig_preset = "COSMIC"
beam = False
```

Variables are defined in a configuration schema written in YAML data serialization language. Different types of *Run*s have different configuration schemes.

### 2.2 Notification queue

Unlike DCS software, where the current state of all the subsystems is requested, our DAQ meta-software assume that it will be notified by subsystems when monitored variables change their values. Each notification contains a timestamp and at least one variable-value pair. All notifications are gathered in an ordered queue. The queue is periodically flushed into a persistent storage.

Having records of all the changes in the values, we can find out in what state the system was at any time slot. If some values were unknown for a certain period of time, all the data for all time slots during that period may be considered suspicious or invalid.

The notifications can be sent to the queue by any application via web-API, ØMQ[2] endpoint or its TCP-gate. Any application, such as alarm system daemon or a live dashboard user interface, can subscribe to the new messages in the queue via ØMQ Pub/Sub or HTTP Server-Sent Events API.

---

2  ØMQ is a brokerless asynchronous messaging library.

The queue is constantly monitored by the `faqmon` unix daemon and it "pauses" the *Run* when there is a mismatch between the expected and actual variable values. In fact, it creates a pause object in the runs database without interrupting the data acquisition. Subsequently, experts can revise the pauses and delete them if the deviation was acceptable, so the skipped data will be included in the *Run*.

## 2.3 Data processing monitor

This component aims to provide a simplified model of the data processing chain. For each *Run*, the processing sequence is specified as a directed acyclic graph, defined as a list of adjacency tuples. In typical cases, there is a chain of decoding, merging, event building, and filtering.

Each processing instance has to report a unique ID of each processed chunk of data. When the processing of the chunk is finished (successfully or with an error), the final processing status is recorded in the chunk database.

## 2.4 Storage management

This component is intended to avoid "data swamp" formation (practice of creating poorly governed data storages). It can be seen as a virtual directory, or as a simplified version of the Rucio software framework for distributed data storage [3].

This component defines Storage Pools, Directories and Entries objects. Each object is identified by a unique ID, which is a combination of timestamp, source ID, and a random number. The actual filesystems are mapped to the virtual Directory objects, and actual files are represented by Entries. All operations with virtual objects trigger helper scripts that do the same with the real filesystem objects. Support for various file systems is accomplished by creating custom helper scripts for each type of a storage pool.

## 3 Results and plans

ηDAQ is our first attempt to implement the DAQ meta-software concept. In 2019, most of its components were used in the production during the beam tests of SPASCHARM [4] fixed-target experiment at IHEP, Protvino. In 2020, we are going to complete the documentation and make the first public release [5].

## Summary

In this work, we proposed a novel concept "DAQ meta-software," which could potentially be considered as an alternative to DAQ frameworks. The idea is to combine all the common high-level functions of DAQ software in one package. Data processing functionality or any other experiment-specific logic must be implemented separately, and will communicate with the meta-software only via asynchronous notifications and requests. The DAQ chain of the entire experiment is described by the meta-software configuration, which makes it possible to use it in a variety of experiments without making changes to the source code. Our implementation of this approach, ηDAQ, has been tested in production. The first public release will happen soon.